\documentclass[12pt]{article}
\usepackage{amsfonts}
\usepackage{amssymb}
\usepackage{graphicx}
\usepackage{amsmath,graphicx,amsfonts}
\usepackage{epsfig}
\usepackage{subfigure}
\usepackage{dcolumn}

\def\siz{\small}
\def\be{\begin{equation}}
\def\ee{\end{equation}}

\begin{document}
\title{Can cosmic acceleration be caused by exotic massless particles?}
\author{P.C. Stichel$^{1)}$  and W.J.
Zakrzewski$^{2)}$
\\
\siz
$^{1)}$An der Krebskuhle 21, D-33619 Bielefeld, Germany \\ \siz
e-mail:peter@physik.uni-bielefeld.de
\\ \\ \siz
$^{2)}$Department of Mathematical Sciences, University of Durham, \\
\siz Durham DH1 3LE, UK \\ \siz
 e-mail: W.J.Zakrzewski@durham.ac.uk
 }

\date{}
\maketitle

\begin{abstract}

To describe dark energy we introduce a fluid model with no free parameter on the microscopic level.
The constituents of this fluid are massless particles which are a dynamical realisation of the 
unextended $D=(3+1)$ Galilei algebra. These particles are exotic as they live in an enlarged
phase space. Their only interaction is with gravity. A minimal coupling to the gravitational
field, satisfying Einstein's equivalence principle, leads to a dynamically active gravitational 
mass density of either sign. A two-component model containing matter (baryonic and dark)
and dark energy leads, through the cosmological principle, to Friedmann-like equations.
Their solutions show a deceleration phase for the early universe and an acceleration phase for the late
universe. We predict the Hubble parameter $H(z)/H_0$ and the deceleration parameter $q(z)$ and 
compare them with available experimental data.
We also discuss a reduced model (one component dark sector) and the inclusion of radiation.
Our model shows no stationary modification of Newton's gravitational potential.


\end{abstract}

\section{Introduction}
Astrophysical observations (supernovae data \cite{one},\cite{two}) suggest that the universe is undergoing an accelerated
expansion. This conclusion was drawn by interpreting these data
in the framework of the cosmological Friedmann equations which describe the universe as being 
homogeneous and isotropic on the largest scales (cp \cite{four}). Within this framework the origin of the cosmic
acceleration is attributed to an exotic component, called dark energy, which is the source
of repulsive gravitation (due to its negative pressure - according to the present 
interpretation).

But there exist other interpretations of the astrophysical data which do not invoke dark energy:
\begin{itemize}
\item
Cosmic acceleration could be an apparent effect due to the averaging of large scale inhomogeneities in the 
universe (see \cite{five} and the literature quoted therein. For a non-expert explanation 
see \cite{extraa}). However, it is an open question as to
whether this interpretation is in an agreement  with all available cosmological data (see \cite{six}, section 5.3).
\item 
Modification of the geometric part of the Einstein-Hilbert action by replacing the Ricci scalar
$R$ by an arbitrary function of it $f(R)$ or by introducing higher-order derivative terms (see \cite{seven}
and the literature quoted therein). Some models are based on modified teleparallel gravity (see \cite{eight}).
All such models, however, suffer from having to rely on an arbitrary function which cannot be derived 
from more fundamental assumptions.
\end{itemize}

Hence we assume that some sort of dark energy is the cause of the cosmic acceleration. Before we present
our model let us give a very brief critical overview of the presently available dark energy models
(see also \cite{nine}).

The simplest model, also called $\Lambda$CDM model, (see any review of dark energy, {\it eg} \cite{six}) involves the use of a positive cosmological constant $\Lambda$  whose value has to 
be determined from experimental data. Its small value (as determined by such considerations) causes some
problems when we interpret $\Lambda$ as the energy density of the vacuum (cp \cite{eleven}, \cite{twelve}). The most popular
dynamical dark energy models use instead a scalar field (see the reviews in \cite{eleven}). However, such models
have less predictive power as one can always construct a scalar field potential that gives rise to 
a given cosmic evolution \cite{nine}.

Another class of models unifies dark matter and dark energy into a one component dark sector. Then the acceleration  comes
from a new kind of interaction within the dark sector. In the case of a Chaplygin gas this interaction is
given by an ad hoc assumed equation of state with negative pressure (see \cite{twelve} and the literature mentioned therein).
Other models use a complex scalar field (see \cite{extra} and the literature cited therein) 
or a phenomenological antifriction force which can be understood as a non-minimal coupling
of the cosmic gas to the curvature \cite{thirteen}.

In summary; so far we do not have any dark energy model which has been derived from fundamental physics \cite{six}.
All known models contain at least one new parameter in the microscopic action \cite{fourteen}.

In this paper we introduce our dark energy model which, on the microscopic level, contains no new
parameters. To do this we start with the well known fact that cosmology can be discussed without
using general relativity as the basic Friedmann equations can be derived from the Newtonian 
gravity (cp. \cite{Extra}). 
If we now want to consider some new nonrelativistic particles as the cause of cosmic acceleration they
must necessarily be massless as the usual massive particles always lead to attractive gravitation.
The possibility of having nonrelativistic massless particles as a dynamical realisation of an extended 
Galilei algebra has already been discussed in some of our recent papers \cite{fifteen}. In the present 
paper we show that massless particles can exist also as a dynamical realisation of an unextended Galilei 
algebra (a related realisation for massless particles moving with infinite velocity has quite recently been found by Duval and Horvathy \cite{Duval}).

The existence of nonrelativistic massless particles may appear strange at first sight; however, as we show in 
section 2, such particles possess a modified relation between energy and momentum (or velocity) and so
they live in an enlarged phase space.  For this reason we will call these particles `exotic'.
Due to the enlarged phase space we have some freedom on how to introduce the gravitational 
coupling for our particles. Here we will do this in a minimal way which satisfies the general form of the 
Einstein equivalence principle but which does not use the concept of a rest mass of the gravitating 
particle. This can be stated in the form of the requirement that ``a freely falling observer does not feel 
any effect of gravitation" \cite{sixteen}. This minimal coupling has the important property that, in a many 
exotic particle system, it leads to a dynamically generated active gravitational mass density of either sign which
can then be a source of the gravitational field. Then a fluid mechanical generalisation of this model can serve 
as a new model for dark energy.

A further extension, to a two-fluid model, including baryonic- and dark matter besides dark energy, then leads,
using the cosmological principle, to Friedmann-like equations for the cosmological scale factor $a(t)$.
The solutions show a deceleration phase for the early universe and an acceleration phase for the late universe.

Furthermore, we show that our model allows for a one component description of the dark sector on large
({\it ie} cosmological) scales. The choice
between these two possibilities has to be made by comparison with the observational data on galactic scales.

We have also looked at the influence of our dark energy sector on local systems. We show that,
in particular, 
it does not lead to the modification of Newton's gravitational potential.



The paper is organised as follows. In section 2 we present our nonrelativistic massless particle model coupled
minimally to gravitation. In section 3 this model is generalised further and extended to a two component
fluid model for matter (baryonic and dark one) and dark energy.  In section 4 we show that we do not run into
instabilities with our model.
In section 5 we describe some solutions, which satisfy the cosmological 
principle,  of the corresponding fluid dynamical equations. 
These solutions show a decelerating universe at early times and an accelerating one at late times.
We will see that on the cosmological level our model is an effective one component 
model for the dark sector. In section 6 we include radiation. 
Observational consequences discussed in section 7 are:
\begin{itemize}
\item The prediction of the Hubble parameter $H(z)/H_0$ and of the deceleration parameter $q(z)$, having fixed two integration constants by using  measured cosmological parameters.
\item The proof that Newton's gravitational potential requires no stationary modification.
\end{itemize}

Some technical details are given in appendix A. In appendix B we speculate on a relativistic generalisation of our nonrelativistic particles describes tachyons. We conclude with some final remarks (section 8).

\section{Nonrelativistic massless particles and their gravitational interaction}
In our second paper in \cite{fifteen} we have introduced the Lagrangian
\begin{equation}
\label{eone}
L\,=\,p_i(\dot x_i- y_i)\,+\,q_i\dot y_i\,-\,\frac{1}{2\kappa}q_i^2,
\end{equation}
where, in the three dimensional case, $x_i$ ($y_i$) are the components of spatial position (velocity) of a particle
and $p_i$ ($q_i$) 
are the components of the corresponding momenta. We use Euclidean metric and Einstein's summation convention with $i=1,2,3.$ An overdot represents a time derivative.

The Lagrangian (\ref{eone}) leads to a dynamical realisation of the acceleration-extended Galilei group in any dimension with one central 
charge ($\kappa$) for a non-interactive massless particle. Without the last term in (\ref{eone}) we have a dynamical
realisation of the Galilei group without any central charge ({\it ie} without any free parameter).

To show that we note that when $\kappa=\infty$, the equations of motion that follow from (\ref{eone}) are
\begin{equation}
\label{etwo}
\dot x_i=y_i,\quad \dot p_i=0,\quad \dot q_i=-p_i,\quad \dot y_i=0.
\end{equation}

These equations correspond to the canonical Poisson brackets (PBs)
\begin{equation}
\label{ethree}
\{x_i,\,p_j\}\,=\,\delta_{ij},\qquad \{y_i,\,q_j\}\,=\,\delta_{ij},
\end{equation}
which can be derived from the Hamiltonian 
\be
\label{ham}
H\,=\,p_iy_i.
\ee

If we now introduce the conserved  Galilean boost generator $K_i$ which is given by
\begin{equation}
\label{efour}
K_i\,=\,p_it\,+\,q_i
\end{equation}
we find that
\begin{equation}
\label{efive}
\{p_i,\,K_j\}\,=\,0,
\end{equation}
which clearly shows that we are dealing with a massless particle.

Going the other way round, {\it ie} by starting with $m=0$, with (\ref{efive}) as a requirement,
it can be shown that the Lagrangian
\begin{equation}
\label{enewseven}
L\,=\,p_i(\dot x_i-y_i)\,+\,q_i\dot y_i,
\end{equation}
defined in a 12-dimensional phase space, is the minimal one \cite{newstichel}.

Furthermore, we note that the conserved angular momentum is given by
\be
\label{ang}
\vec J\,=\,\vec x\times \vec p\,+\,\vec y\times \vec q
\ee 
and that the Poisson brackets of $\vec p$, $\vec K$, $\vec J$ and $H$ build the unextended 
Galilei algebra. The presence of the second term in (\ref{ang}) shows that
our particles possess a nontrivial spin.

To couple this particle to gravity we start with the general form of Einstein's equivalence 
principle. In a nonrelativistic context this can be stated as follows: locally, {\it ie} at each fixed
space point $\vec x$, a gravitational force $-\vec \bigtriangledown\phi (\vec x,t)$ is equivalent
to a time-dependent acceleration $\vec b(t)$. The only known equation of motion for the 
particle trajectory $\vec x(t)$ satisfying this form of the equivalence principle is given by the Newton 
law:
\begin{equation}
\label{esix}
\ddot x_i(t)\,=\,-\partial_i \phi(\vec{x}(t),t)
\end{equation}
because (\ref{esix}) is invariant with respect to arbitrary time-dependent 
translations (cp \cite{seventeen})
\begin{equation}
\label{eseven}
x_i\,\rightarrow\, x_i'\,=\,x_i\,+\,a_i(t)
\end{equation}
if $\phi(\vec{x},t)$ transforms to
\begin{equation}
\label{eeight}
\phi'(\vec{x}\,',t)\,=\,\phi(\vec{x},t)\,-\,\ddot a_i(t)\,x_i\,+\,h(t).
\end{equation}

Hence considering $\phi(\vec{x},t)$ as an external gravitational field we can take its
interaction term with our particle $L_{int}$ in the form:
\begin{equation}
\label{enine}
L_{int}\,=\,q_i\,\partial_i\,\phi(\vec{x}(t),t)
\end{equation}
Clearly, with this term, the equation of motion for $x_i$ is given by (\ref{esix}) and the second equation in 
(\ref{etwo}) becomes $\dot p_i=q_k\partial_k\partial_i\phi$. Then
our system is invariant 
under arbitrary time-dependent translations (\ref{eseven}) where $\phi$ transforms
according to (\ref{eeight})  with $q_i$ and $p_i$ being invariant.

\section{Two-fluid dynamics}
In this section we consider a two-fluid cosmological model where one fluid component  $M$ consists of massive
matter (baryonic and dark one) and the other fluid $D$ consists of the exotic massless particles, introduced
in the previous section and representing dark energy. The only interaction considered within the fluids 
and between them is gravitational.

\subsection{Lagrange picture}
First we generalise the dark energy model introduced in the previous section to the continuum case by 
introducing comoving coordinates $\vec{\xi}\in R^3$ \cite{eeightteen}, add continuous massive matter
with its standard gravitational interaction and use the usual Lagrangian for the gravitational field.

Then our Lagrangian is given by
\begin{equation}
\label{eten}
L\,=\,L_M\,+\,L_D\,+\,L_{\phi},
\end{equation}
where
\begin{equation}
\label{eeleven}
L_M\,=\,m\int \,d^3\xi\,\left(y_i^M(\dot x_i^M\,-\,\frac{1}{2}y_i^M)\,-\,\phi(\vec{x}^M,t)\right),
\end{equation}
where $m$ is a mass parameter giving (\ref{eeleven}) the correct dimension,
\begin{equation}
\label{etwelve}
L_D\,=\,\int\,d^3\xi\,\left(p_i(\dot x_i^D-y_i^D)\,+\,q_i^D\dot y_i^D\,+\,q_i\,\partial_i\,\phi(\vec{x}^D,t)\right)
\end{equation}
and
\begin{equation}
\label{ethirteen}
L_{\phi}\,=\,-\frac{1}{8\pi G}\,\int\,d^3x\,\left(\vec{\nabla}\phi(\vec{x},t)\right)^2.
\end{equation}
In these expressions all phase space variables are functions of $\vec{\xi}$ and $t$, {\it ie} $\vec{x}^M=\vec{x}^M(\vec{\xi},t)$ etc.

Note that both $L_D$ and $L_M$ are invariant, up to a total time derivative, under the transformations (\ref{eseven}-\ref{eeight}).

The equations of motion (EOM) corresponding to $L$ are given by
\begin{itemize}
\item $M$ sector
$$  \dot x_i^M\,=\,y_i^M$$
\begin{equation}
\label{efourteen}
\dot y_i^M\,=\,-\partial_i\,\phi(\vec{x}^M,t)
\end{equation}

\item $D$ sector
$$  \dot x_i^D\,=\,y_i^D$$
\begin{equation}
\label{efifteen}
 \dot q_i^D\,=\,-p^D_i
\end{equation}
$$ \dot y_i^D\,=\,-\partial_i\,\phi(\vec{x}^D,t)$$
$$  \dot p_i^D\,=\,q_k \partial_k\partial_i\,\phi(\vec{x}^D,t)$$
\item $\phi$ sector
\end{itemize}
\begin{equation}
\label{esixteen}
 \triangle \phi(\vec{x},t)\,=\,4\pi G \int \,d^3\xi\left(
m\delta(\vec{x}-\vec{x}^M(\vec{\xi},t))\,+\,q_i(\vec{\xi},t)\,\partial_i\,\delta(\vec{x}-
\vec{x}^D(\vec{\xi},t))\right).
\end{equation}
The last term in (\ref{esixteen}) represents a dynamically generated active gravitational 
mass density.

\subsection{Eulerian picture}

In the Eulerian picture the dynamics of the fluid is described in terms of $\vec x$ and $t$ 
dependent fields: particle number density $n(\vec x,t)$, velocity $u_i(\vec x,t)$,
momentum $p_i(\vec x,t)$ and pseudo-momentum $q_i(\vec x,t)$.


Assuming uniform distribution in $\vec \xi$ the Lagrangian phase space variables are
transformed to the Eulerian fields by
\be
n(\vec x,t)\,=\, \int\, d^3\xi\,\delta^3(\vec x-\vec x(\vec \xi,t))
\label{ex}
\ee
and
\be
n(\vec x,t)\,p_i(\vec x,t)\,=\,\int\,d^3\xi\,p_i(\vec \xi,t)\,\delta^3(\vec x-\vec x(\vec \xi,t))
\label{ex1}
\ee
and an analogous expression for $u_i(\vec x,t)$ (in the expression above replace
$p_i(\vec x,t)$ by $u_i(\vec x,t)$ and $p_i(\vec \xi,t)$ by $y_i(\vec \xi,t)$). Similarily 
for $q_i(\vec x,t)$. 
In fact, (\ref{ex1}) holds for any function of relevant variables.

To derive the EOM in the Eulerian picture we follow the standard procedure (cp. \cite{eeightteen})
and obtain from (\ref{efourteen}-\ref{esixteen}) by using
(\ref{ex},\ref{ex1}) the corresponding equations in the 
Eulerian picture:
\begin{equation}
\partial_t\,n^A(\vec x,t)\,+\, \partial_k(n^A u_k^A)(\vec x,t)\,=\,0, \label{a}
\ee
where $A=(M,D)$, 
{\it ie} the continuity equations for the particle number densities $n^M$ and $n^D$,
and from (\ref{esixteen}) the Poisson equation for the gravitational field
\be 
 \triangle \phi(\vec{x},t)\,=\,4\pi G \left(\rho^M\,+\,\partial_i(n^Dq_i)\right),
\label{b}
\ee
where the mass density $\rho^M$ is defined by $\rho^M:=m n^M$.

Note that the last term in (\ref{b}) represents the dynamically generated active gravitational mass density of the 
dark-energy fluid.

We have, in addition, the following Euler equations:
\be
\label{c}
D_t^M\,u_i^M\,=\,-\partial_i\phi
\ee
(from the second equation in (\ref{efourteen})) and from the third equation in
(\ref{efifteen})
$$ D_t^D\,u_i^D\,=\,-\partial_i\phi,$$
where we have defined $D_t^A=\partial_t + u_i^A\partial_i$.

Suppose now that $u_i^M$ and $u_i^D$ obey the same initial conditions. Then 
(\ref{c}) shows that $u_i^D=u_i^M=u_i$ {\it ie} (\ref{c}) becomes one universal
Euler equation valid for all fluid components.
\be
\label{d}
D_t\,u_i\,=\,-\partial_i\phi.
\ee

Finally, the second and fourth equations in (\ref{efifteen})  give
\be
D_t\,q_i\,=\,-p_i,\qquad D_t\,p_i\,=\,q_k\partial_i\partial_k\,\phi.
\label{e}
\ee
Looking at (\ref{d},\ref{e}) we note that, in contrast to standard fluid mechanics,
the two vector fields $\vec p(\vec x,t)$ and $\vec u(\vec x,t)$ are not parallel to each other.

\subsection{Symmetries}

First we note that all our EOM (\ref{a}-\ref{e}) are obviously rotationally
symmetric.

To consider other symmetries we observe that if we perform an infinitesimal time dependent 
translation $\delta x_i=a_i(t)$ we see that
$$ \delta u_i(\vec{x},t) \,=\,\dot a_i(t)\,-\,a_k(t)\,\partial_k\,u_i(\vec{x},t)$$
\begin{equation}
\label{f}
\delta \phi(\vec{x},t)\,=\,-\ddot a_i(t)x_i\,+\,h(t)\,-\,a_k(t)\,\partial_k\,\phi(\vec{x},t)
\end{equation}
and
$$ \delta \zeta (\vec{x},t)\,=\,-a_k(t)\,\partial_k\,\zeta(\vec{x},t)$$
where $\zeta\in(n^A,p_i,q_i)$.
Thus the EOM are invariant under such translations and so, locally, the general form of Einstein's principle
of equivalence is satisfied  as in General Relativity. Moreover, as shown recently 
by one of us (PCS), we obtain, when neglecting the massive matter part, as symmetry
algebra, the expansion-less conformal Galilei algebra with dynamical exponent $z=\frac{5}{3}$ \cite{newstichel}.

\subsection{Stress tensor and pressure}
To see that our massless particle fluid may, indeed, represent dark energy we show now that the pressure can be negative.
To do this we consider the local momentum conservation
\begin{equation}
\partial_t P_i(\vec x,t)\,+\,\partial_j\,T_{ij}(\vec x,t)\,=\,0,
\label{F1}
\end{equation}
where, in our case, the momentum density $P_i$ and stress tensor $T_{ij}$ are given by
\begin{equation}
\label{F2}
P_i(\vec x,t)\,=\,(np_i)(\vec x,t),\quad T_{ij}(\vec x,t)\,=\,(P_iu_j)(\vec x,t)\,+\,{\cal P}\delta_{ij}
\end{equation}
and $\cal{P}$ is the pressure.

Note that the stress tensor is not symmetric. The deeper reason for that is the presence of a spin part in the
conserved angular momentum (see (\ref{ang}) and cp. \cite{Extra4}).

To have the system as simple as possible we consider first a one-dimensional self-gravitating
massless particle fluid. The EOM for the momentum field $p(x,t)$ then, due to the Poisson equation (\ref{esixteen}),
becomes
\begin{equation}
\label{F3}
D_t\,p\,=\,4\pi\,G\,q\,\partial_x(nq).
\end{equation}
Using the continuity equation (\ref{a}), together with (\ref{F3}) we obtain for (\ref{F1}) 
\begin{equation}
\label{F4}
\partial_t \,P\,+\,\partial_x\left(Pu\,-\,2\pi G(nq^2)\right)\,=\,0
\end{equation}
{\it ie} we get the negative pressure
\begin{equation}
\label{F5}
{\cal P}\,=\,-\,2\pi G\,(nq)^2.
\end{equation}

For a three dimensional case we obtain a result similar to (\ref{F5}) if we assume that the 
vector field $n\vec q$ is irrotational (see section 5), {\it ie}
\begin{equation}
\label{F6}
(nq_k)(\vec x,t)\,=\,\partial_k h(\vec x,t).
\end{equation}
Then, from (\ref{esixteen}), we get
\begin{equation}
\label{F7}
\partial_i\phi\,=\,4\pi\,G\,n\,q_i,
\end{equation}
leading to the EOM for the momentum field $p_i$
\begin{equation}
\label{F8}
D_t\,p_i\,=\,4\pi\,G\,q_k\,\partial_i(nq_k)
\end{equation}
and finally, analogously to (\ref{F4}), to the pressure
\begin{equation}
\label{F9}
{\cal P}\,=\,-2\pi\,G\,(nq_k)^2.
\end{equation}
If $n\vec q$ is not irrotational then the stress tensor $T_{ij}$ is more complicated
than that given in (\ref{F2}).

\subsection{Nature of the gravitational mass}

According to the Poisson equation (\ref{b}) the active graviational mass density of the 
dark energy fluid $\hat \rho^D$ is given by
\begin{equation}
\label{eea}
\hat \rho^D\,=\,\partial_i(nq_i).
\end{equation}
This result should be compared with the expression,  from general relativity, for the 
perfect fluid
\begin{equation}
\label{eeb}
\hat \rho^D\,=\,\left( \rho^D\,+\,\frac{3{\cal P}}{c^2}\right),
\end{equation}
where $c^2\rho^D$ is the energy density in the rest frame of the fluid and ${\cal P}$
- the corresponding pressure. Neglecting the massive matter component and considering the particular case of an irrotational field $n\vec q$ (see (\ref{F6})) we find from 
the Lagrangian (\ref{etwelve}-\ref{ethirteen}) and the transformations (\ref{ex}-\ref{ex1})
that the energy density ${\cal E}^D(\vec x,t)$ is given by 
\begin{equation}
\label{eec}
{\cal E}^D\,=\,n\,p_i\,u_i\,-\,2\pi\,G(nq_i)^2
\end{equation}
and so
\begin{equation}
\label{eed}
\rho^D\,=\,-\frac{2\pi G(nq_i)^2}{c^2}.
\end{equation}

Then, with ${\cal P}$ given by (\ref{F9}), we obtain for (\ref{eeb})
\begin{equation}
\label{eef}
\hat \rho^D\,=\,-\frac{8\pi G(nq_i)^2}{c^2}.
\end{equation}

The reason for the difference between (\ref{eef}) and (\ref{eea}) is obvious: in our
nonrelativistic massless particle model there is no place for the velocity 
of light $c$! Furthermore, we note that (\ref{eea}) is not built from the 
components of the energy-stress tensor and, in fact, (\ref{eea}) is the simplest
Galilei invariant expression in our model with the dimension of mass density.

\section{Stability considerations}
The Hamiltonian corresponding to the Lagrangian (\ref{etwelve}) of the dark energy fluid
component is linear in the momenta $p_i^D$ and thus not bounded from below. This property 
is well known for any higher-derivative Lagrangian (theorem of Ostrogradsky \cite{final2})
and it arises in our case because our one-particle Lagrangian considered in section 2 may be
understood as the limiting case of the higher-order Lagrangian (\ref{eone}), which in the 
configuration space takes the form $L=\frac{\kappa}{2} \ddot x_i^2$ (see  \cite{fifteen}$_2$). However, for a free particle, this does not concern us too much, as the 
Hamiltonian (\ref{ham}) can always be transformed to a positive quadratic form by a complex-valued canonical transformation\footnote{Similar ideas can be used to demonstrate the absence of ghosts in the  Pais-Uhlenbeck model. This was recently shown, in a different way, by Bender and Mannheim \cite{Bender}}:
$$ p_i\,=\,iap_i'\,-\, bq_i',\quad y_i\,=\,-bq_i'\,-\,iap_i',$$
\begin{equation}
\label{E0}
q_i\,=\,\frac{1}{2b}y_i'\,-\, \frac{i}{2a}x_i',\quad x_i\,=\,-\frac{i}{2a}x_i'\,-\,
\frac{1}{2b}y_i',
\end{equation}
where $a,b$ are arbitrary real numbers,
leading to
\begin{equation}
\label{E00}
H\,=\,a^2p_i^{,2}\,+\,b^2q_i^{,2}.
\end{equation}
Thus all serious problems (like, {\it eg} the collapse in the classical case or a nonunitary 
time-development in the quantum case) may arise only in the presence of interactions
(cp \cite{final3}, \cite{final4}).

Another possible instability of interacting theories containing negative energy involves the spontaneous  decay of any state into a collection of positive and negative energy particles
(cp. \cite{final4}). This instability is excluded in our model due to the particle
number conservation (\ref{a}).

To find out what happens in our fluid model it is sufficient to consider only a self-gravitating massless particle system, {\it ie} to neglect its dark matter component.
To treat things as simple as possible, in the following we consider only a one-dimensional
system. In subsection 4.1 we show that a two-particle system at zero energy can, indeed,
collapse. When, in the next subsection,  we generalise these considerations to the continuum,
{\it ie} a fluid dynamical case, we find that in this case the collapse does not take 
place. This reassures us in our belief that the three dimensional case is also collapse free.

\subsection{Two-particle case}

Specialising the EOM (\ref{efifteen}), for $d=1$, to the two-particle case we obtain for the relative motion:
\begin{equation}
\label{E1}
\ddot x(t)\,=\, 4\pi\,G\,q(t)\,\delta(x(t))
\end{equation}
and
\begin{equation}
\label{E2}
\ddot q(t)\,=\,2\pi\,G\,q^2(t)\,\delta'(x(t))
\end{equation}
where we have defined the relative variables (the indices 1,2 label the two particles)
$$ x:=\, x_1\,-\,x_2,\quad q:=\,q_1\,-\,q_2. $$
The variables of the two-particle centre, $R:=\frac{1}{2}(x_1+x_2)$ and $Q:=\frac{1}{2}(q_1+q_2)$
satisfy
\begin{equation}
\label{E3}
\ddot Q\,=\,0,\quad \hbox{and}\quad \ddot R\,=\,-4\pi\,G\,Q\,\delta(x).
\end{equation}
 To obtain these equations we have used the fact that, for a generic $x\in R^1$, $\partial_x\phi(x,t)$ is given, due to (\ref{esixteen}), by
\begin{equation}
\label{E4}
\partial_x\,\phi(x,t)\,=\,4\pi\,G\, \sum_{\alpha=1}^2\,\delta(x-x_{\alpha}(t))q_{\alpha}(t);
\end{equation}
- here we have taken the 
particular solution $Q(t)=0$ of (\ref{E3}).

Then the energy of the two-particle system is given by
\begin{equation}
\label{E5}
E\,=\,-\frac{\dot q\dot x}{2}\,+\,\pi\,G\,q^2\,\delta(x).
\end{equation}

A solution of the EOM (\ref{E1}), (\ref{E2}), for vanishing energy $E$, is clearly given by
\begin{equation}
\label{E6}
\dot x\,=\,2\pi\,G\,\lambda\,q^2,\qquad \dot q\,=\,\lambda^{-1}\,\delta(x),
\end{equation}
where $\lambda$ is arbitrary and needed for dimensional reasons.

If we now take $\lambda>0$ so that
\begin{equation}
\label{E61}
\dot x(t)\ge 0,\quad \hbox{for \ all}\quad t\in R^1
\end{equation}
and choose
\begin{equation}
\label{E62}
x(t)<0,\, \dot x(t)>0 \quad \hbox{for}\quad t<t_0\quad \hbox{and}\quad x(t_0)=0
\end{equation}
we obtain from (\ref{E6}) that
\begin{equation}
\label{E7}
q(t)\,=\,q_0(1-\theta(t-t_0))
\end{equation}
and
\begin{equation}
\label{E8}
\dot x(t)\,=\,\begin{cases}{2\pi \,G\, \lambda\,q_0^2\quad \hbox{for \ } t<t_0}\\
                    {0\quad\quad\quad\quad \hbox{for \ } t>t_0}\end{cases}
                    \end{equation}
 {\it ie} the two particles collide at $t=t_0$ and stay together for all later times.
 This is a collapse situation.

 \subsection{Hydrodynamic case}
 
 In the Eulerian picture for $d=1$ the analogue of (\ref{E4}) is now
 \begin{equation}
 \label{E9}
 \partial_x\,\phi\,=\,4\pi\,G\,n\,q
 \end{equation}
 and so the hydrodynamic EOM obtained from (\ref{a}), (\ref{d}) and (\ref{e})
 take the form
 \begin{equation}
 \label{E10}
 \partial_t\,n\,+\, \partial_x(un)\,=\,0,
 \end{equation}
 \begin{equation}
 \label{E11}
 D_t\,u \,=\,-4\pi\,G\,n\,q
 \end{equation}
 and
 \begin{equation}
 \label{E12}
 D_t^2\,q\,=\,-4\pi\,G\,q\,\partial_x(nq).
 \end{equation}
 
 In analogy to (\ref{eec})  
     the energy density ${\cal E}(x,t)$ is given by
 \begin{equation}
 \label{E13}
 {\cal E}\,=\,n\left(-(D_t\,q)u\,-\,2\pi Gnq^2\right).
 \end{equation}
 
 Next we proceed as in the two particle case.
 We make the ansatz
 \begin{equation}
 \label{E14}
 u\,=\,2\pi\,G\,\lambda\,q^2.
 \end{equation}
 Then from (\ref{E11}) we find that
 \begin{equation}
 \label{E15}
 D_t\,q\,=\,-\lambda^{-1}\,n
 \end{equation}
 which, together with (\ref{E14}), demonstrates the vanishing of the energy density $\cal E$.
 
 Then, as can be easily checked, the unique solution of the remaining EOM (\ref{E10}),
 (\ref{E12}) and (\ref{E14}) is given by
 \begin{equation}
 \label{E16}
 n(x,t)\,=\,n_0(=\hbox{const}),\qquad q(x,t)\,=\,\frac{-n_0\,t\,+\,c}{\lambda}
 \end{equation}
 demonstrating a collapse-free situation.

\section{Cosmological solutions of fluid dynamics equations}

In order for the universe to be homogeneous and isotropic on large scales we require, as usual, that
\be
\label{g}
n^A\,\,=\,n^A(t)
\ee
and
\be
\label{h}
u_i\,=\,\frac{\dot a(t)}{a(t)}\,x_i,
\ee
where $a(t)$ is the cosmic scale factor.

Then (\ref{d}) tells us that
\be
\label{i}
\partial_i \,\phi\,=\,x_i\,\varphi(t)
\ee
with 
$$ \varphi(t)\,=\,-\frac{\ddot a}{a}.$$
 
Putting (\ref{h}) and (\ref{i}) into the second equation in (\ref{e}) gives us
\be
\label{j}
D_t\,p_i\,=\,-q_i\,\frac{\ddot a}{a}.
\ee

To solve the first equation in (\ref{e}) and (\ref{j}) we make an ansatz
\be
\label{k}
q_i\,=\,f_q(t)\,x_i,\quad \hbox{and}\quad p_i\,=\,f_p(t)\,x_i.
\ee

Then, using (\ref{h}-\ref{j}), we eliminate $f_p$ and get
\be
\label{l}
\ddot f_q\,+\,2\,\frac{\dot a}{a}\,\dot f_q\,=\,0,
\ee
which can be integrated once giving us
\be
\label{m}
\dot f_q(t)\,=\,\frac{\beta}{a^2(t)}\, \quad \hbox{with}\quad \beta=\hbox{const.}
\ee
Furthermore, with (\ref{g}) and (\ref{h}) the continuity equations (\ref{a}) can be integrated as usual 
giving us for $\rho^M$ and $n^D$
\be
\label{n}
\rho^M\,=\,\frac{M}{\frac{4\pi}{3}\,a^3(t)} 
\ee
and
\be
\label{aaaaa}
n^D\,=\,\frac{D}{\frac{4\pi}{3}\,a^3(t)},
\ee
where $M$ and $D$ are positive constants. 

Inserting (\ref{i}) and (\ref{n}) into the Poisson equation (\ref{b}) we get from (\ref{k}) 
\be
\label{o}
-\ddot a\,=\,\frac{G}{a^2}\,(M+3Df_q),
\ee
 where $f_q$ should be taken as a solution of (\ref{m}).
Eq. (\ref{o}) is one of our Friedmann-like equations.

We should now distinguish two cases:
\begin{itemize}
\item  $\beta=0$ which implies $f_q$=const. Then eq.(\ref{o}) gives us that
\be
\label{p}
\ddot a>0\quad \hbox{for}\,\,\hbox{any}\,\,t\quad \hbox{if}\quad f_q<-\frac{M}{3D}
\ee
{\it ie} we obtain an accelerated expansion for all times (this contradicts the known cosmological facts).
\item
$\beta\ne0$. Then 
putting (\ref{m}) into (\ref{o}) we get
\be
\label{q}
-\ddot a\,=\,\frac{\dot f_q G}{\beta}(M+3Df_q).
\ee
Integrating once we find
\be
\label{qq}
-\dot a\,=\,\frac{f_q G}{\beta}(M+\frac{3}{2}Df_q)\,+\,c_1
\ee
Multiplying (\ref{qq}) by $\dot f_q$ and using (\ref{m}) on the l.h.s  we obtain
\be
\label{r}
-\frac{\dot a \beta}{a^2}\,=\,\frac{\dot f_q f_q G}{\beta}(M+\frac{3}{2}Df_q)\,+\,c_1\dot f_q
\ee
which after integration gives us
\be
\label{s}
\frac{\beta}{a}\,=\,\frac{G}{2\beta}\,f_q^2\,(M+Df_q)\,+\,c_1f_q\,+\,c_0,
\ee
\end{itemize}
where $c_0$ and $c_1$ are integration constants.


Let us now discuss, given (\ref{s}), the behaviour of $f_q$ as a function of the scale factor $a$.

Performing the transformation:
\be
\label{ss}
f_q\,\rightarrow\,\,g(a):=\,f_q\,+\,\frac{M}{3D}
\ee
we arrive at (redefining $c_0$ and $c_1$)
\be
\label{t}
g^3(a)\,+\,c_1g(a)\,+\,c_0\left(1\,-\,\frac{a_t}{a}\right)\,=\,0,
\ee
where we have defined 
\be
\label{tt}
a_t\,:=\,\frac{2\beta^2}{GDc_0}.
\ee

Let us look now at the solution of ({\ref{t}) with the constants $c_0$ and $c_1$ being positive, $c_{0,1}>0$.
First of all we note that the scale factor $a$ may serve as a measure of time due to $\dot a>0$ (expanding universe).

For $a<a_t$ we have $g(a)>0$ and so, due to (\ref{o}) $\ddot a<0$.
So, for $a<a_t$, we are in the deceleration phase of the early universe. On the other hand, clearly,
for $a>a_t$ we have $g(a)<0$ and then, due to (\ref{o}), $\ddot a>0$. So, for $a>a_t$ 
we are in the acceleration phase of the late universe and we see that $a_t$ defines the transitional
scale factor at which the deceleration stops and the acceleration takes over.  
It can easily be seen that the condition $c_{0,1}>0$ is also necessary to obtain these results.

Next we observe that by differentiating (\ref{t}) with respect to $a$ we have
\be
\label{u}
g'(a)\,=\,-\frac{a_tc_0}{a^2(c_1+3g^2(a))},
\ee
where $'$ denotes the derivative with respect to $a$.
If we now put (\ref{u}) into (\ref{m}) we get
\be
\label{uu}
\dot a\,=\,-\frac{\beta(c_1+3g^2(a))}{c_0a_t}
\ee
thus showing that, for $\dot a>0$, we need $\beta<0$.

Eq. (\ref{uu}) is our second Friedmann-like equation. Note that the first Friedman-like
equation (\ref{o}) is a consequence of the second one (\ref{uu}) if $g(a)$ is a solution of the cubic 
equation (\ref{t}).

To integrate (\ref{uu}) we need the explicit form of $g(a)$.
To obtain $g(a)$ we note that $g(a)$ is the real valued solution of the cubic equation (\ref{t}). This solution 
is given by 
\be
\label{aaa}
g(a)\,=\,u_+(a)\,+\,u_-(a)
\ee
with
\be
\label{uuu}
u_{\pm}(a)\,=\,\left(-\frac{q}{2}\,\pm\,\left[\left(\frac{c_1}{3}\right)^3\,+\,\left(\frac{q}{2}\right)^2\right]^{\frac{1}{2}}\right)^{\frac{1}{3}},
\ee
where
$$q:=\,c_0\left(1\,-\,\frac{a_t}{a}\right).$$

Then, from (\ref{uu}) we find that
\be
\label{v}
t-t_0\,=\,\frac{c_0\,a_t}{\vert \beta\vert}\,\int\,da\,\frac{1}{c_1\,+\,3g^2(a)},
\ee
with $g(a)$ given by (\ref{aaa}). 

In the Appendix A we present a detailed discussion of the evaluation of (\ref{v}) in terms of roots of 
(\ref{t}). As our final results are not very transparent let us mention here some asymptotic results:
\begin{itemize}
\item At large $a$, {\it ie} $a\gg a_t$,  $a$ grows linearly with $t$. This follows from the observation that at large $a$, 
$q$ goes to $c_0$ and so the integrand in (\ref{v}) becomes independent of $a$.
\item At $a$ very close to $a_t$ we get from (\ref{t}) that
\be
\label{newa}
g(a)\,\simeq \,-\frac{c_0}{c_1\,a_t}\,(a-a_t).
\ee
Then, by choosing $t_0$ as the time at which $a=a_t$ we obtain from (\ref{v}) 
\be
\label{newb}
t-t_0\,\simeq \frac{-i}{\delta}\,\log\,\frac{1+i\gamma (a(t)-a_t)}{1-i\gamma (a(t)-a_t)},
\ee
where $\gamma:=\frac{\sqrt{3}c_0}{c_1^{\frac{3}{2}}a_t}$ and $\delta:=\frac{2\sqrt{3}\vert \beta\vert}{c_1^{\frac{1}{2}}a_t^2}.$

Inverting (\ref{newb}) and taking the first terms of the power series expansion in $t-t_0$ we obtain
\be
\label{newc}
a(t)-a_t\,=\,\frac{c_1\vert \beta\vert}{c_0a_t}(t-t_0)\,+\,\frac{\vert \beta\vert^3}{c_0a_t^5}(t-t_0)^3\,
+\,O((t-t_0)^5).
\ee

Considering (\ref{newa}) with (\ref{o}) and (\ref{u}) it is easy to see that higher order corrections to 
(\ref{newa}) do not change the first two terms in the expansion (\ref{newc}).
\item For small $a$ {\it ie} for $a \ll a_t$ we obtain from (\ref{t}) 
\be
\label{newd}
g(a)\, \simeq \left(\frac{c_0a_t}{a}\right)^{\frac{1}{3}}
\ee
leading to, due to (\ref{v}) with $t_0=0$ and $a(0)=0$, 
\be
\label{newe}
a(t)\,\sim t^{\frac{3}{5}}
\ee
thus showing that the combined effect of matter and dark energy at the early times differs from the  behaviour of the matter dominated universe for which $a(t)\sim t^{\frac{2}{3}}$.
\end{itemize}

Note that this result (\ref{newe}) is exactly the scale invariant solution for $a(t)$ corresponding to the dynamical exponent $z=5/3$ \cite{newstichel}.

\subsection{Dark sector with one or two components?}

In section 3 we made the usual assumption that the dark sector possesses a two-component 
structure. However, our results for the Friedmann-like equations (\ref{o}), (\ref{uu}) and 
for the equation (\ref{t}) determining $g(a)$ are all independent of the constant $M$, 
defined by (\ref{n}). Thus, as long as we do not compare (\ref{o}) and (\ref{uu}) with
the original Friedmann equations which would be physically senseless due to the different
nature of the gravitational mass in our model (see section 3.5), it is sufficient to keep 
the dark energy fluid (now to be called ``dark fluid'') as the only component within the dark 
sector. In our case, this dark fluid takes over the role of dark matter and dark energy,
at least on large scales, like in the cases of the Chaplygin gas \cite{twelve} or the complex scalar field \cite{extra}. The baryonic component, which corresponds to about 4\% of the energy of the universe, is negligible on these scales.

To describe the universe correctly, at the scale of galaxies, the dark fluid must behave like 
dark matter; {\it ie} exhibit attractive gravitation at local scales (see \cite{extra} and the literature
cited therein). This point still has to be examined in more detail.

\section{Cosmology including radiation}

Including radiation (photons), and also massless neutrinos, within our framework, would require a full 
relativistic treatment. However, what we really need here is somewhat less ambitious. For the cosmology as outlined above we need a description of radiation as a nonrelativistic fluid\footnote{Note that within a hydrodynamic description of radiation the velocity field at a point $\vec x$ 
is an average over all directions of radiation velocities whose modulus is therefore less than $c$ (it might even be small when compared to $c$).} component $R$ with an equation of state parameter (defined as the ratio of pressure and energy density) \cite{six}
\be
\label{newf}
\omega^R\,=\,\frac{1}{3}.\ee
 
To get the required result we follow McCrea \cite{Extra1} and Harrison \cite{Extra2} who extended Newtonian 
cosmology by taking pressure into account. For a homogeneous and isotropic universe we have therefore
to add to our hydrodynamic equations the continuity equation for the radiation energy density $c^2 \rho^R$
\be
\label{newg}
\dot{{\rho}}^R\,+\,4\frac{\dot a}{a}\,\rho^R\,=\,0,
\ee
whose solution is given by
\be
\label{newh}
 \rho^R\,=\,\frac{R}{\frac{4\pi}{3}a^4(t)},
\ee
where $R$ is a positive constant.
Furthermore we must change the Poisson equation (\ref{b}) by adding to its right hand side the active gravitational radiation mass density $2 \rho^R(t)$ leading to
\be
\label{newi}
\triangle \phi\,=\,4\pi G(\rho^M\,+\,\partial_i(n^D q_i)\,+\,2\rho^R).
\ee

From (\ref{newi}) we conclude that the first Friedmann-like equation (\ref{o}) now becomes
\be
\label{newj}
-\ddot a\,=\,\frac{G}{a^2}\,(M\,+\,3Df_q\,+\,\frac{2R}{a}).
\ee

Unfortunately, when $R\ne 0$,
 it is not possible to integrate analytically the coupled system of differential equations (\ref{newj}) and (\ref{m}).
Nevertheless, we can conclude, as usual, that at very early times the last term in (\ref{newj}) 
dominates, {\it ie} the universe is radiation dominated. In the following, we will consider, 
as we have already done in section 5, the universe only for the later times, {\it ie} when 
the last term in (\ref{newj}) is negligible.

\section{Observational consequences}
Our exotic massless particles possess no non-gravitational interaction, neither with
the particles of the Standard Model nor with the dark matter particles. Thus their 
existence can only lead to observational consequences at cosmological scales (see sections 7.1-2) and,
perhaps, also at local scales (see section 7.3).


\subsection{Predicting the Hubble and the deceleration parameters from our model} 

To calculate the Hubble parameter $H$ in our model we introduce the redshift  $z$ by 
\be 
\label{w1}
a\,=\,\frac{1}{1+z}
\ee
and then consider
\be
\label{w2}
H(z):\,=\,\frac{\dot a}{a}(z).
\ee
Then from (\ref{o}) we find that $H(z)$ is given by
\be
\label{w3}
H(z)\,=\,\frac{\vert \beta\vert c_1}{ac_0a_t}\,\left(1\,+\,3\frac{g^2(a)}{c_1}\right)\left|_{a=\frac{1}{1+z}}\right ..
\ee

Next we define $H_0$ as $H_0:=H(z=0)$ and so find that
\be
\label{w4}
h(z):\,=\,\frac{H(z)}{H_0}\,=\,\frac{1\,+\,3\frac{g^2(\frac{1}{1+z})}{c_1}}{1\,+\,3\frac{g^2(1)}{c_1}}(1+z).
\ee

In a similar way we see that the deceleration parameter $q(z)$ defined as
$$ q(z)\,=\,-\frac{\ddot a}{a H^2(z)}$$ is given by
\be
\label{w5}
q(z)\,=\,\frac{6a_t}{a}\frac{c_0}{c_1^{\frac{3}{2}}}\frac{g(a)}{c_1^\frac{1}{2}}\,
\left(1\,+\,3\frac{g^2(a)}{c_1}\right)^{-2}\left|_{a=\frac{1}{1+z}}\right .
\ee

Note that both are functions of only $a_t$ and of $\kappa=\frac{c_0}{c_1^{\frac{3}{2}}}.$ Hence to determine 
them we need two experimental data. 

Before we try to determine $q(z)$ and $H(z)$ let us observe that 
there are a few things we can say about their behaviour for any values of the two parameters.
First of all, we easily see  from (\ref{newd}) that at large $z$,
{\it ie} for $z\gg z_t=\frac{1-a_t}{a_t}$, 
\begin{equation}
\label{Z1}
\frac{g\left(\frac{1}{1+z}\right)}{\sqrt{c_1}}\,\simeq\,\left(\frac{z}{1+z_t}\right)^{\frac{1}{3}}\,
\left(\frac{c_0}{c_1^{\frac{3}{2}}}\right)^{\frac{1}{3}}
\end{equation}
and so from (\ref{w5}) we get that, for $z\gg z_t$,  
\begin{equation}
\label{Z2}
q(z)\,\simeq  \frac{2}{3}.
\end{equation}

Moreover, $h(z)$ is monotonically increasing. To see this we take (\ref{w3}) and note that 
\begin{equation}
\label{Z3}
h(z)\,=\,k\left( 1+3\tilde g^2(z)\right)(1+z),
\end{equation}
where we have defined
$$\tilde g(z):=c_1^{-\frac{1}{2}}\,g(\frac{1}{1+z})$$
and, similarily, the overall positive constant $k$.
Then (\ref{u}) is equivalent to
\be
\label{Z4}
\tilde g'(z)\,=\,\frac{\kappa}{(1+z_t)(1+3\tilde g^2)}\,>\,0.
\ee

Then 
\be
\label{Z5}
h'(z)\,=\,k\left(1+3\tilde g^2\,+\,\frac{6\kappa (1+z)\tilde g}{(1+z_t)(1+3\tilde g^2)}\right)\,>\,0.
\ee

In addition, from (\ref{Z1}) and (\ref{Z3}) we see that, for $z\gg z_t$,
\be
\label{Z6}
h(z)\,\sim\,z^{\frac{5}{3}}.
\ee

\subsection{Estimation of $H(z)$ and of $q(z)$}

To obtain our `predictions' for $h(z)$ and $q(z)$ we use the data from the first reference in 
\cite{data}. They give us $q(0)=-0.57$ and $z_t=0.71$, both with small errors which we do not mention here
as the curves we will show here depend very little on the exact values of these parameters.
These values are obtained by fitting the matter part $\Omega_m$ of the $\Lambda CDM$ model
to observational data. $q(0)$ then serves to determine the constant $\kappa$ in our model.
We will give the curves obtained with these values subscript $S$.

We can also use the model independent values from the other two references in \cite{data}. The data from the paper by
Cunha are $q(0)=-0.73$ and $z_t=0.49$ and from the paper by Lu et al $q(0)=-0.788$ and $z_t=0.632$.
The curves corresponding to them will carry the indices $C$ and $L$ respectively. Note that the values
of $\kappa$ for the three cases are $\kappa= 0.8667$ (S), 0.977 (C) and 1.166 (L).

In fig. 1 and 2  we plot our predictions for $H(z)$ and $q(z)$ respectively.
We have normalised $H(z)$ to its value at $z=0$, so in fact, our plots are of $h(z)$
In fig. 3 we present the corresponding values for $g(z)$.
We note that all 3 cases are quite similar.

We have also attempted to compare our results to the experimental data given in
table 1 in  \cite{data1}. The results given there have large experimental errors
and are given only for a few values of $z$. Hence they will
not be too conclusive or reliable.  However, to perform any comparison we need
the value of $H_0$. We can, of course,  take this value from \cite{two}$_2$. There we find 
$H_0=70.5$.
 Hence in fig 4 we present our data (with the normalisation fixed by $H_0=70.5$)
and compare them with the experimental data (constructed from the data in \cite{data1})
corresponding to the experimental data + 1 standard deviation error (called 'maximum')
and -1 standard deviation error (called 'minimum').
We note the general agreement and so we are heartened by this result.


\begin{figure}
\begin{center}
\includegraphics[angle=270, width=10cm]{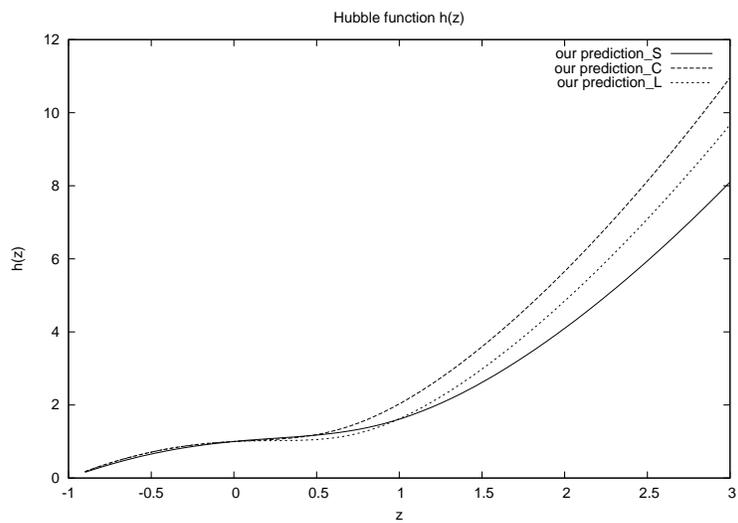}
\caption{Our prediction for $h(z)$}
\end{center}
\end{figure}

\begin{figure}
\begin{center}
\includegraphics[angle=270, width=10cm]{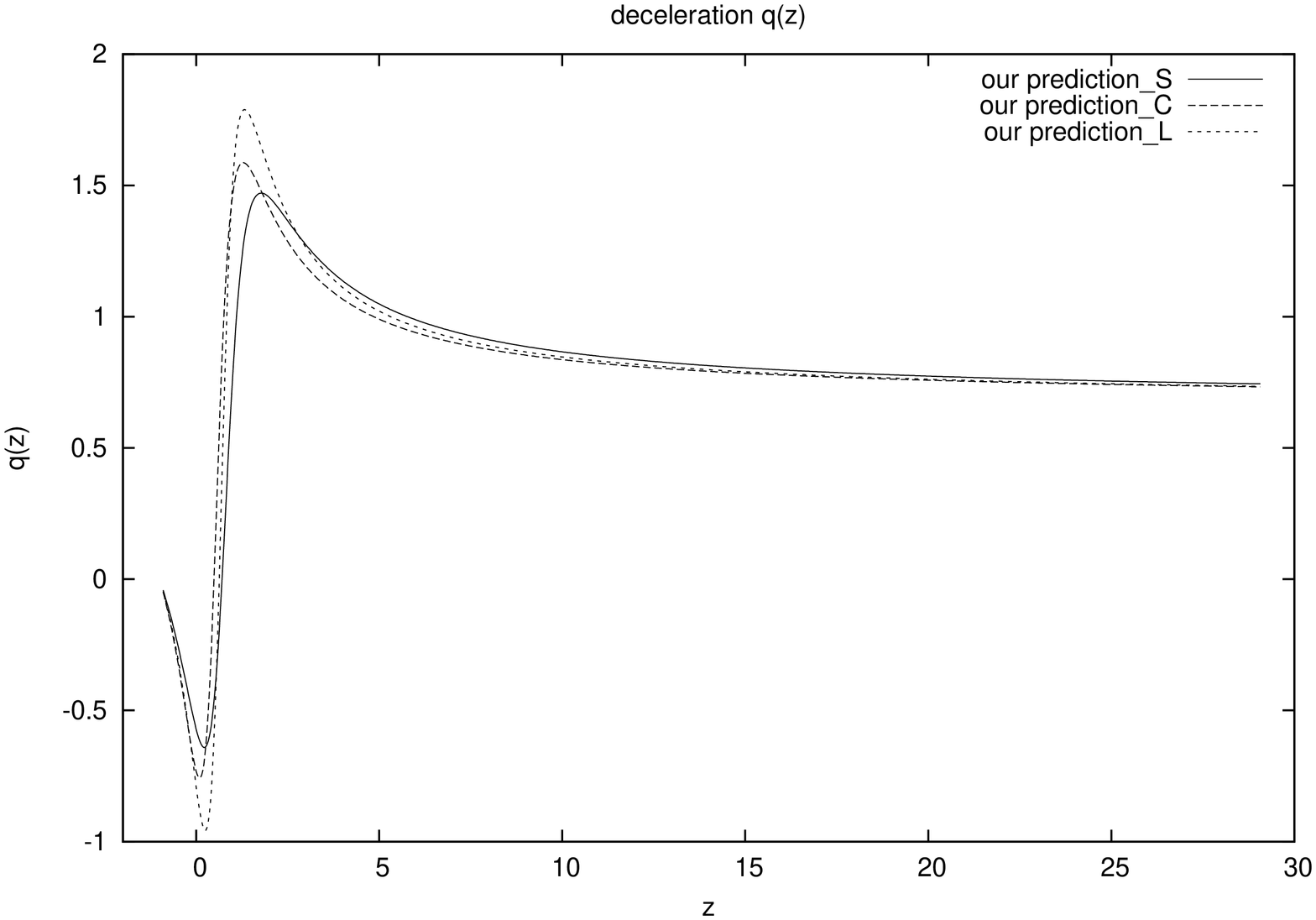}
\caption{Our prediction for $q(z)$}
\end{center}
\end{figure}


\begin{figure}
\begin{center}
\includegraphics[angle=270, width=10cm]{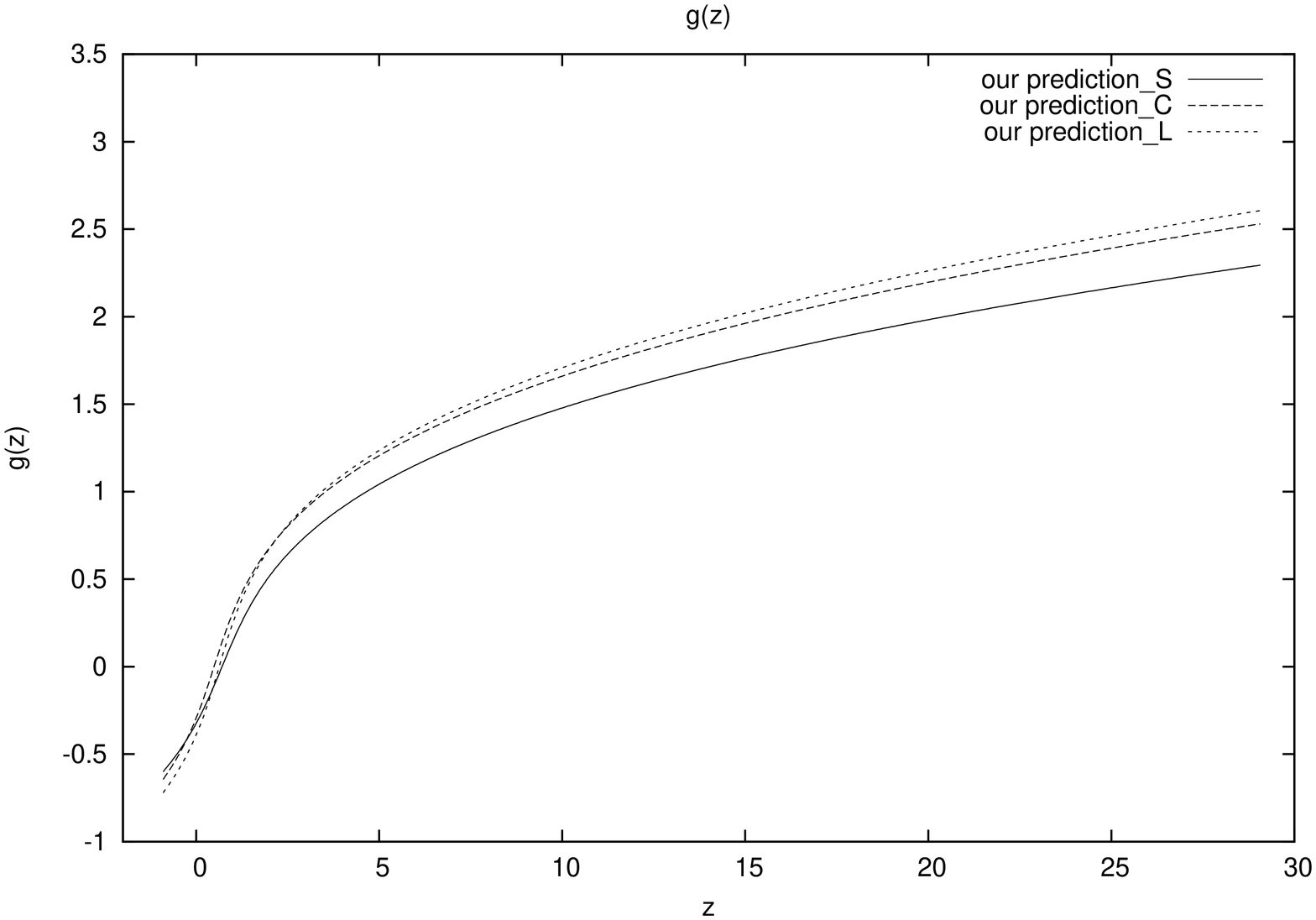}
\caption{Our prediction for $g(z)$}
\end{center}
\end{figure}

\begin{figure}
\begin{center}
\includegraphics[angle=270, width=9cm]{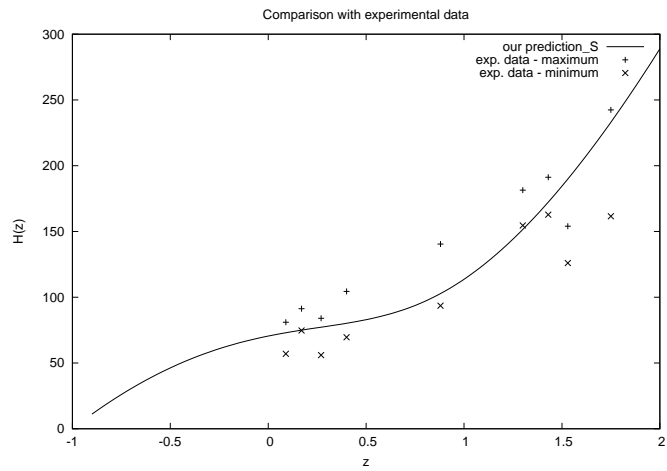}
\caption{Comparison with experimental data;$H(z)$ in units $km \, sec^{-1}\, Mpc ^{-1}$  }
\end{center}
\end{figure}

\subsection{Influence on local systems}
Here we look at the problem of how a two-body system, bound by the standard Newtonian potential, may be affected
by the dark sector proposed in this paper. To study this we consider two different mechanisms:
\begin{itemize}
\item
The effect of the dark fluid at cosmological scales giving rise to an additional time-dependent term for the 
two body potential
\be
\label{a1}
\delta \phi(r,t)\,=\,-\frac{r^2}{2}\,\frac{\ddot a}{a}.
\ee
The equations for the two-body relative motion then takes the form:
\be
\label{a2}
\ddot{\vec{r}}\,=\,\frac{\ddot a}{a}\,\vec{r}\,-\,\frac{G\mu}{r^3}\vec{r},
\ee
where $\mu$ is the reduced mass. 

As we do not have the explicit form of the time dependence of the scale factor $a(t)$ we use instead $a$ as
a measure of time. Then (\ref{a2}) leads to the following differential equation for
$\vec{r}(a)$:
\be
\label{a3}
\mathop{r}^{\rightarrow''}\,\dot{a}^2\,+\,\mathop{r}^{\rightarrow'}\,\ddot a\,=\,\frac{\ddot a}{a}\vec{r}\,-\,\frac{G\mu}{r^3}\vec{r}
\ee
or using the Friedman-like equations (\ref{o}) and (\ref{uu}), we obtain
\be
\label{a4}
\mathop{r}^{\rightarrow''}(a)\beta^2\frac{(c_1+3g^2(a))^2}{c_0^2\,a_t^2}\,-\,\frac{3DG}{a^2}g(a)\mathop{r}^{\rightarrow'}(a)\,=\,
\ee
$$
-\frac{3DG}{a^3}g(a)\vec{r}(a)\,-\,\frac{G\mu}{r^3(a)}\vec{r}(a),
$$
where $g(a)$ is given by (\ref{aaa}) and a prime denotes differentiation with respect to $a$.

To solve (\ref{a4}) numerically we would have to know
besides the constants $a_t$ and $c_0/c_1^{\frac{3}{2}}$ known from 7.1, also the values of constants $\beta$ and $c_0$.
 Recent estimates
of the effects caused by $\delta \phi(r,t)$ in the case of a constant $w^D<-1$ \cite{twentyone} have found
observable effects on a time-scale given by billions of years\footnote{For consideration of more general
astronomical structures see \cite{twentytwo} and the literature cited therein.}. We expect similar
results for our model.
\item 
The other issue involves a possible modification of Newton's gravitational
potential by a local, stationary dark energy fluid. To study this we consider a point mass $m$ located at 
$\vec{x}=0$. We will show that the corresponding stationary dark energy flow leads to a vanishing extra 
gravitational mass density $\partial_i(\rho^Dq^D_i)$ and so there is no extra contribution
to $\phi(r)$.
To see this we consider the $D$ sector of our equations of motion given in subsection 3.2 for the stationary 
case. They become
\be
\label{a5}
\partial_k(n^Du_k)\,=\,0,
\ee
\be
\label{a6}
u_k\,\partial_ku_i\,=\,-\partial_i\phi
\ee
\be
\label{a7}
u_k\,\partial_k\,q_i^D\,=\,-p_i^D
\ee
\be
\label{a8}
u_k\,\partial_k\,p_i^D\,=\,q_k^D\,\partial_k\partial_i\phi
\ee
together with the Poisson equation
\be
\label{a9}
\triangle \phi\,=\,4\pi\,G(m\delta(\vec{x})\,+\,\partial_i(n^Dq_i^D)).
\ee

Then we use (\ref{a7}) and (\ref{a6}) to eliminate $p_i^D$ and $\partial_i\phi$ in (\ref{a8}) 
and obtain
\be
\label{a10}
u_k\,\partial_k\,u_l\,\partial_l\,q_i^D\,=\,q^D_k\,\partial_k\,u_l\,\partial_l\,u_i.
\ee
Looking at (\ref{a10}) we note that it implies that $q^D_i$ has to be proportional to $u_i$ 
\be
\label{a11}
q_i^D\,\sim\,u_i
\ee
and so, due to (\ref{a5}), we obtained the desired result {\it ie}
\be
\label{a12}
\partial_k\,(n^Dq_k^D)\,=\,0.
\ee
\end{itemize}

\section{Final Remarks} 
Given that there are already many dark energy models, what are the reasons why we have introduced a further one? The reasons are twofold:
\begin{itemize}
\item There are no free parameters in the 
microscopic formulation of our model. 
\item Our model introduces new physical ideas in the form of nonrelativistic massless particles whose minimal
coupling to gravity leads to the generation of an active gravitational mass density of either sign.
\end{itemize}

This last point poses the question about the relation of these new physical ideas to Newton's and Einstein's
theory of gravity. As our particles are a dynamical realisation of the unextended Galilei algebra they fit into the general scheme of nonrelativistic physics. The gravitational coupling, satisfying Einstein's equivalence
principle, leads to the same equation of motion (8) in configuration space as in the massive case.
Thus we can consider our model of a gravitationally coupled, nonrelativistic massless particle as an extension of Newton's theory of gravity. 

However, it seems to be not possible to obtain our model as a nonrelativistic limit of a relativistic model. Massless relativistic particle models possess conformal Poincar\'e symmetry leading, in the nonrelativistic limit, to conformal symmetry \cite{fifteen}$_1$,
{\it ie} $z=1$. In our case, for a one-component dark sector, $z=\frac{5}{3}$. In Appendix B
we have speculated that the relativistic generalisation of our Galilean massless particles
are tachyons. However, it may be that we are here in a situation similar to Ho\v{r}ava gravity \cite{finala}; {\it ie} we have nonrelativistic symmetry in the ultraviolet limit (small $t$) and approach General Relativity only in the infared (large $t$) limit. However, to have such
a picture we may have to modify our model. This is a challenge for further research.

As a drawback of our model one can consider the existence of additional dimensions in phase space.
However, such a case is already well known from the related case of nonrelativistic massless fields
(Galilean electromagnetism) in which the lagrangian formulation requires the introduction of auxiliary fields
\cite{final}. In our case the additional degrees of freedom lead in the Friedmann-like equations 
to undetermined constants which are integration constants along the additional phase space dimensions.
The question then arises as to whether these constants can be determined a priori by some physical 
arguments. This point is currently under investigation.


\section{\bf Appendix A}

Here we demonstrate that the integral (\ref{v}) can be calculated in a closed form.

First we note that due to (\ref{u}) we have
\begin{equation}
\label{A5}
c_0a_t\,\int \,da\,(c_1+3g^2(a))^{-1}\,=\,-\int\,da\,a^2\,g'(a).
\ee

Next we change the integration variable $a\rightarrow g(a)$ and use (\ref{t}) to rewrite the right hand side of 
(\ref{A5}) as
\be
\label{A6}
-\,(c_0a_t)^2\,\int\,dg\,(g^3+c_1g+c_0)^{-2}.
\ee

We define the roots of the cubic equation
\be
\label{A2}
g^3\,+\,c_1g\,+\,c_0\,=\,0
\ee
as $g_i$. They are given by 
\be
\label{A3}
g_1\,=\,v_+\,+\,v_-,\quad g_2\,=\,-\frac{v_++v_-}{2}\,+\,\frac{v_+-v_-}{2}i\sqrt{3},\,\quad g_3\,=\,g_2^{\star},
\ee
with
\be
\label{A3}
v_{\pm}\,=\,\left(-\frac{c_0}{2}\,\pm\,\left[\left(\frac{c_1}{3}\right)^3\,+\,\left(\frac{c_0}{2}\right)^2\right]^{\frac{1}{2}}\right)^{\frac{1}{3}}.
\ee
Next we perform the decomposition
\be
\label{A7}
(g^3\,+\,c_1g\,+c_0)^{-1}\,=\,\prod_{i=1}^3 (g-g_i)^{-1}\,=\,\sum_{i=1}^3\,a_i(g-g_i)^{-1},
\ee
where $a_i$ are given by
\be
a_i\,=\,\left((g_i-g_{i+1})(g_i-g_{i-1})\right)^{-1}.
\ee
Here $i=1,2,3$ and cyclic permutation is assumed.

Putting all this together we perform the integration in (\ref{A6}) and obtain
\be
\label{A1}
c_0a_t\,\int \,da\,\frac{1}{c_1+3g^2(a)}\,=\,(c_0a_t)^2\,\sum_{i=1}^3\,a^2_i\frac{1}{g(a)-g_i}
\ee
$$
\qquad -2(c_0a_t)^2\,\sum_{i<j}^3\frac{a_ia_j}{g_i-g_j}\,\log\frac{g(a)-g_i}{g(a)-g_j}.
$$

Clearly 
 $a_1=a_1^{\star}$ and $a_2^{\star}=a_3$.

\section{\bf Appendix B}

Here we discuss a possible relativistic correspondence of the nonrelativistic massless particles
introduced in section 2. Clearly, they cannot correspond to either massive or massless relativistic
particles. However, they could correspond to tachyons which can be seen as follows:
\begin{itemize}
\item The relativistic generalisation of the equations of motion (\ref{etwo}) are given by
the derivatives of the corresponding four-vectors with respect to the relativistic parameter 
$\tau$:
\be
\label{relat}
\dot x_\mu\,=\,y_\mu,\quad \dot p_\mu\,=\,0,\quad \dot q_\mu\,=\,-p_\mu,\quad \dot y_\mu\,=\,0.
\ee
\item From the second and fourth equations in (\ref{relat}) we see that
\be
\label{relattwo}
p_\mu y^{\mu}\,=\,\hbox{const.}
\ee
However, in order to reproduce, in the non-relativistic limit, the energy relation (\ref{ham}) 
the constant appearing on the right hand side of (\ref{relattwo}) must vanish, {\it ie} we must have
\be
\label{relatthree}
p_\mu y^{\mu}\,=\,0.
\ee
\item From (\ref{relatthree}) we see that
\be
\label{relatfour}
p_\mu p^{\mu}\,=\,\frac{(\vec p\cdot \vec v)^2}{c^2}\,-\,\vec p^2\,=\,\le-\left(1\,-\,\frac{v^2}{c^2}\right)\vec p^2\,<\,0.
\ee
\end{itemize}

{\bf Acknowledgments}: We would like to thank Peter Horvathy,  Wolfgang Kundt and  Simon Ross for 
a critical reading of the first version of the manuscript and very helpful comments. We are grateful to the referee
for his suggestions which have led to the substantial improvement of the paper.

\end{document}